\documentclass[twocolumn,prb,showpacs,preprintnumbers,amsmath,amssymb,superscriptaddress]{revtex4}
\usepackage{upgreek}
\usepackage{graphicx}% Include figure files
\usepackage{dcolumn}% Align table columns on decimal point
\usepackage{bm}% bold math
\usepackage{xcolor}
\usepackage{ulem}

\begin{document}

\title{Buried heterostructure vertical-cavity surface-emitting laser with semiconductor mirrors}% Force line breaks with \\

\author{G. Zhao}%
\affiliation{%
CREOL, College of Optics and Photonics, University of Central Florida, Orlando, FL 32816}

\author{Y. Zhang}%
\affiliation{%
CREOL, College of Optics and Photonics, University of Central Florida, Orlando, FL 32816}

\author{D.G. Deppe}%
\email{ddeppe@creol.ucf.edu}
\affiliation{%
CREOL, College of Optics and Photonics, University of Central Florida, Orlando, FL 32816}

\author{K. Konthasinghe}%
\affiliation{%
Dept. of Physics, University of South Florida, Tampa, FL 33620}

\author{A. Muller}%
\email{mullera@usf.edu}
\affiliation{%
Dept. of Physics, University of South Florida, Tampa, FL 33620}

\date{\today}% It is always \today, today,
             %  but any date may be explicitly specified

\begin{abstract}
We report a buried heterostructure vertical-cavity surface-emitting laser fabricated by epitaxial regrowth over an InGaAs quantum well gain medium. The regrowth technique enables microscale lateral confinement that preserves a high cavity quality factor (loaded $Q\approx$ 4000) and eliminates parasitic charging effects found in existing approaches. Under optimal spectral overlap between gain medium and cavity mode (achieved here at $T$ = 40 K) lasing was obtained with an incident optical power as low as $P_{\rm th}$ = 10 mW ($\lambda_{\rm p}$ =  808 nm). The laser linewidth was found to be $\approx$3 GHz at $P_{\rm p}\approx$ 5 $P_{\rm th}$. \end{abstract}

% PACS, the Physics and Astronomy
                             % Classification Scheme.
%\keywords{Suggested keywords}%Use showkeys class option if keyword
                              %display desired
\maketitle

Since their inception, \cite{iga1988ses} vertical-cavity surface-emitting lasers (VCSELs), have dramatically improved in terms of speed,\cite{anan2007pis} efficiency,\cite{amann2012eeh} and output power,\cite{higuchi2012hpd} and are now widely available commercially. One of the principle advantage of VCSELs is a scalable production process, but VCSELs also feature continuous single-mode operation, \cite{huffaker1994nod, ohiso2002sto} mode-hop free tuning, \cite{changhasnain2000tvc} and improved beam quality\cite{gustavsson2006def} compared to edge emitting devices. Nevertheless, reducing VCSEL dimensions into the nanoscale has proved challenging, despite being of great interest for future integration of optics and microelectronics. Three-dimensionally confining heterostructures have demonstrated progress in this direction, including electrically-pumped room-temperature operation,\cite{zhou2000eis}  but under size reduction these structures are expected to suffer parasitic carrier loss associated with etched and exposed interfaces. A fully-{\it buried} heterostructure gain structure has the potential to dramatically improve VCSEL performance by eliminating such parasitic charging effects in the perimeter region of the VCSELÕs gain material. The buried heterostructure (BH) is expected to produce lower threshold, higher efficiency, and higher modulation speed for microscale VCSEL devices. In addition, improved current funneling and heat dissipation in a BH VCSEL may permit operation further above threshold and consequently lead to a further reduction in the laser linewidth.\cite{kim2012nlo} For sensing applications a narrow linewidth is essential for reaching higher resolution. However, because VCSEL performance is also strongly dependent on cavity loss, a BH VCSEL requires a design that can also form a high quality factor microcavity. Although there has been rapid progress in the development of microcavity technology in recent years, \cite{vahala2003om} buried heterostructure designs have typically been non-epitaxial.\cite{ohiso2001bhv}

Here we report a BH quantum well VCSEL in which the epitaxial cavity is overgrown on the active material rather than on the perimeter of the semiconductor cavity.\cite{iga1988ses, changhasnain1993ltb} We describe the design, surface topography, and optical characteristics of devices based on lithographically defined mesas with a diameter as small as 3 $\upmu$m. We find that under optimal spectral overlap lasing is observed with a threshold power of the same order as that of ordinary VCSELs.\cite{disopra1999niv} High-resolution measurements with a scanning Fabry-Perot interferometer reveal a laser linewidth of several GHz just above threshold. 

\begin{figure}[b]
\includegraphics[width=3in]{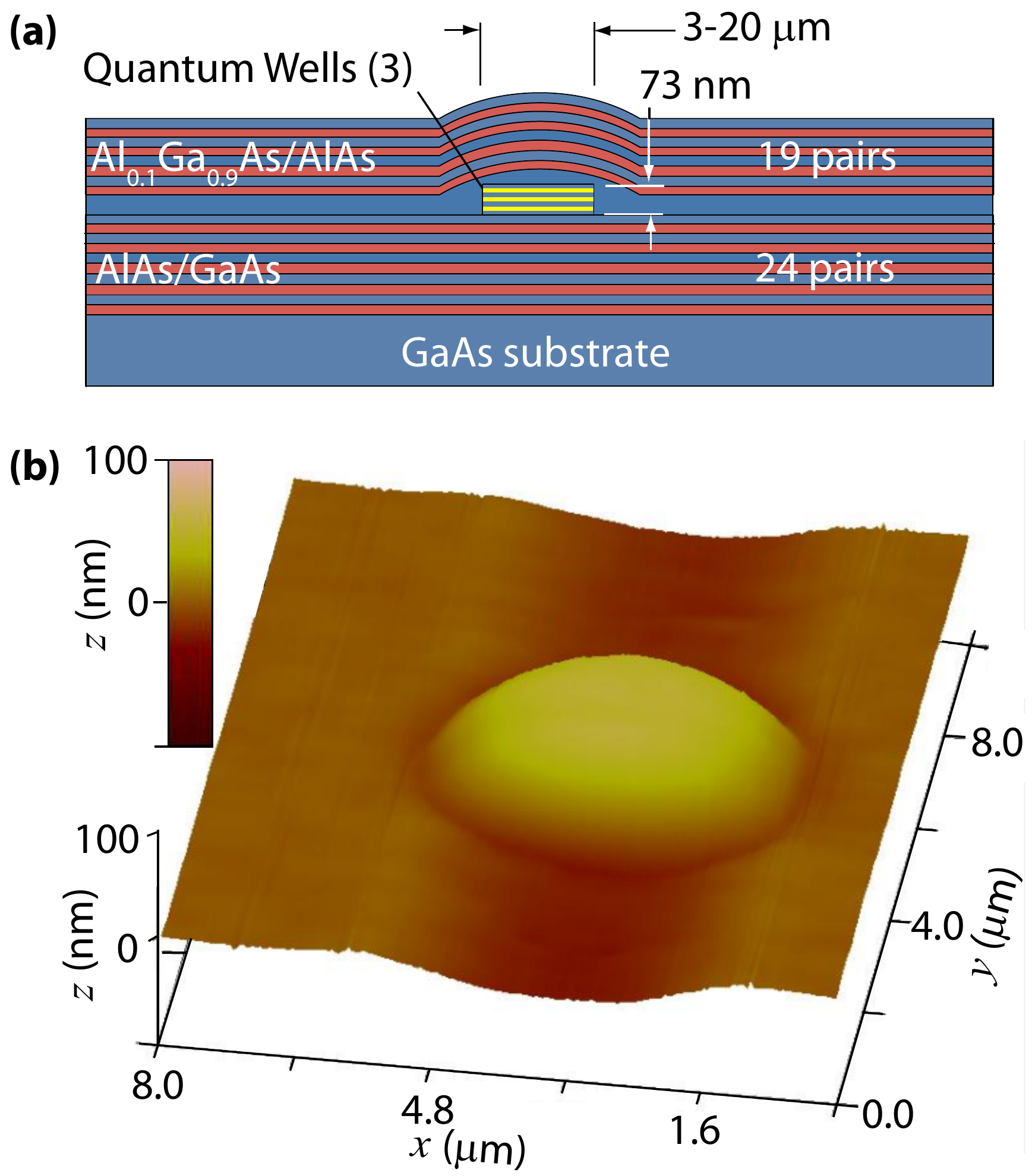}
\caption{\label{schematic} (Color online) (a) Schematic of buried heterostructure vertical-cavity surface-emitting laser. (b) Atomic force microscope image of such as laser for which the lithographically defined mesa had a diameter of 3 $\upmu$m.}
\end{figure}

Our devices were grown by molecular beam epitaxy (MBE) on a GaAs substrate [Fig. 1(a)]. A lower distributed Bragg reflector (DBR) consisting of 24 GaAs/AlAs quarter-wave pairs served as the bottom mirror onto which a gain region of 3 InGaAs quantum wells was grown. Arrays of mesas with diameters ranging from 3 $\upmu$m to 20 $\upmu$m and with a height of 73 nm were defined in this gain region by ultraviolet lithography and subsequent wet etching. A final HF etch was used prior to reloading the sample in the growth system. The top DBR, consisting of 19 Al$_{0.1}$Ga$_{0.9}$As/AlAs quarter-wave pairs, was grown over the etched mesa. The regrowth process is derived from methods developed in prior work using a semiconductor quantum dot gain medium.\cite{lu2005lsa, muller2006hqa}

The surface topography of our overgrown sample was obtained by atomic force microscopy (AFM). As seen in Fig. 1(b) for a cavity based on a 3 $\upmu$m diameter mesa, the overgrown surface assumes the shape of a convex lens, with a height close to that of the mesa from which it originates. The lens in Fig. 1(b) is actually slightly elongated in the $x$ direction with an aspect ratio of 1.3 due to anisotropic epitaxial growth. The cross-section of the lens of Fig. 1(b) along the $y$ direction is shown in Fig. 2(a).

\begin{figure}[t]
\includegraphics[width=3.2in]{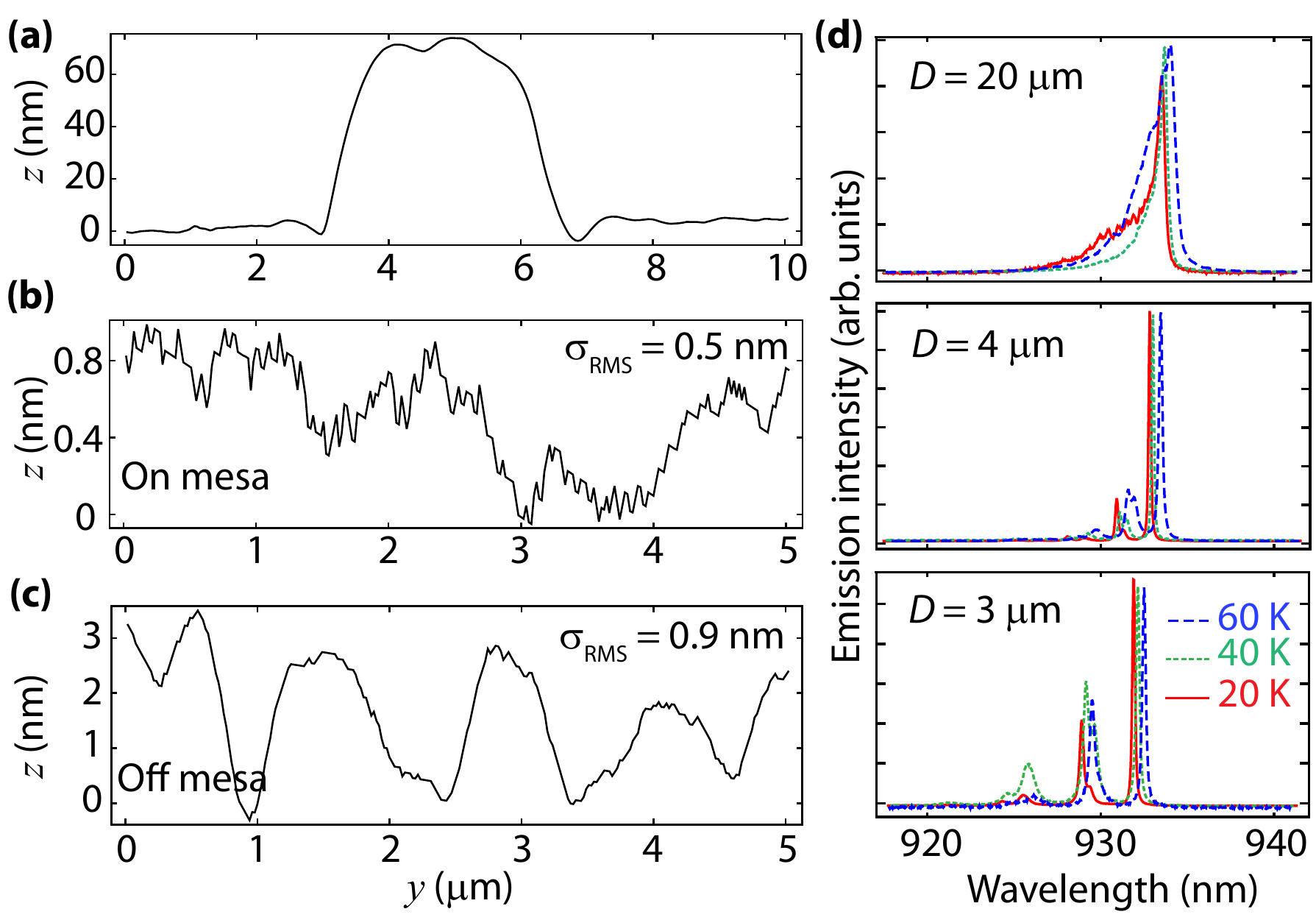}
\caption{\label{xsection} (Color online) (a) AFM linescan of BH VSCEL of Fig. 1(b). (b) AFM linescan near the center of a large lithographically-defined square mesa (80$\times$100 $\upmu$m) (c) Same as in (b) but outside the mesa. (d) Photoluminescence spectra of four different BH VCSELs, fabricated using mesas with diameters of 20 $\upmu$m, 6 $\upmu$m, 4 $\upmu$m, 3 $\upmu$m (top to bottom). The solid red, short-dashed green, and long-dashed lines represent measurements at $T$ = 20 K, $T$ = 40 K, and $T$ = 60 K, respectively.}
\end{figure}

An essential requirement of the regrowth process is a high surface quality so that scattering losses are minimized. A simple estimate of required root-mean-square (RMS) surface microroughness, $\sigma_{\rm RMS}$, can be obtained using Ruze's equation, $S=(4\pi\sigma_{\mathrm{RMS}}/\lambda)^2$, which provides an expression for mirror scattering loss, $S$, at a given wavelength.\cite{ruze1952} For a Fabry-Perot cavity of length $L$, the cavity quality factor, $Q$, is related to the total cavity round trip loss, $\Gamma$, through the expression $\Gamma=2\pi L/\lambda Q$. Therefore, for a cavity that is about one wavelength long, the maximum microroughness that can be tolerated to sustain a quality factor $Q$, is estimated as $\sigma_{\mathrm{RMS}}\approx\lambda/2\sqrt{2\pi Q}$. Thus, in the present case, for $Q\approx$ 5000 we must have $\sigma_{\mathrm{RMS}}\lesssim$ 0.8 nm. AFM linescans in Fig. 2(b) and (c) show a zoomed in measurement on, and off a (80$\times$100 $\upmu$m) mesa location, respectively. The corresponding root-mean-square surface microroughness is $\sigma_{\rm RMS}$ = 0.5 nm and $\sigma_{\rm RMS}$ = 0.9 nm, respectively, revealing a smooth overgrowth compatible with a quality factor on the order of 5000 or larger.

\begin{figure}[b]
\includegraphics[width=3.3in]{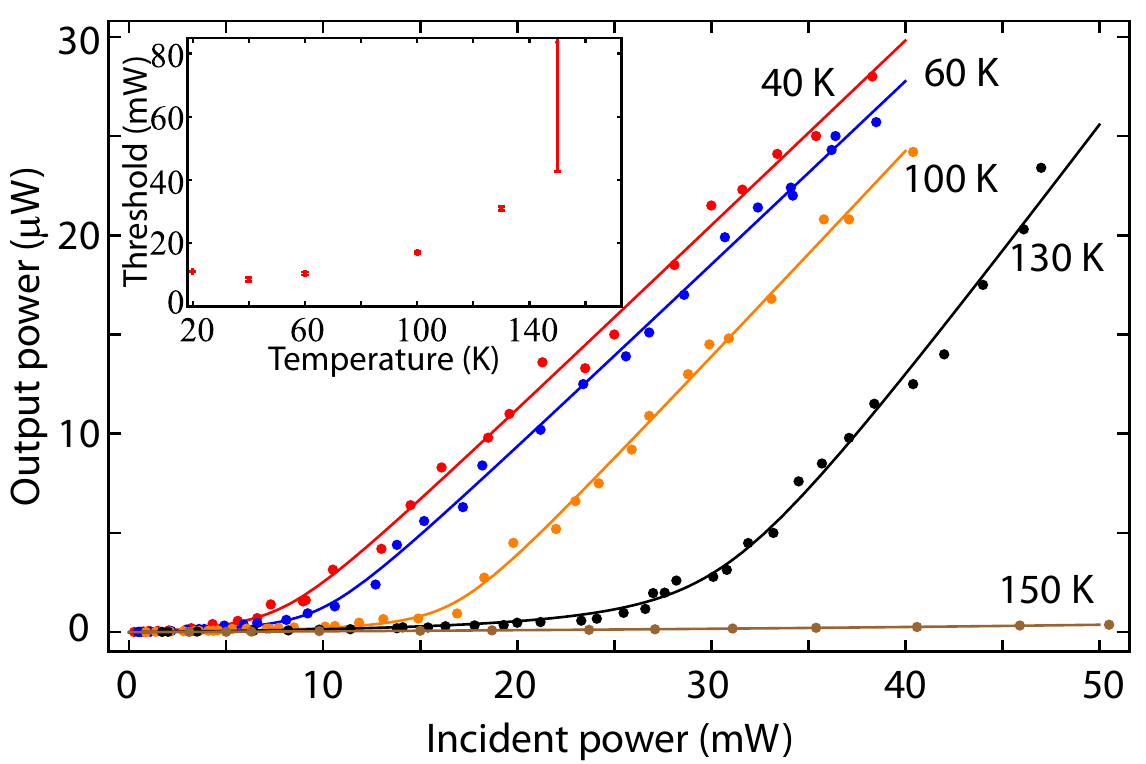}
\caption{\label{lasing} (Color online) Lasing characteristics of a BH VCSEL based on a 3 $\upmu$m diameter mesa. The inset shows the extracted threshold as a function of temperature.}
\end{figure}

Normalized photoluminescence (PL) spectra of structures based on mesas with diameters of 20 $\upmu$m, 4 $\upmu$m, and 3 $\upmu$m are shown in Fig. 2(d). The wavelength of the pump laser was $\lambda_{\rm p}$ = 808 nm. With decreasing mesa diameter the spectra clearly reveal an increase in the spectral separation of the microcavity resonances, accompanied by a blue shift due to the increased lateral confinement.

Laser oscillation was investigated using the same experimental configuration as that for PL, but applying a higher pump power. Figure 3 shows the output power versus incident pump power for a BH VCSEL based on a 3 $\upmu$m mesa at temperatures ranging from 40 K to 150 K. The extracted threshold is plotted in the inset versus temperature, with a minimum threshold measured near $T\approx$ 40 K. This temperature corresponds to the designed value for the spectral gain offset. Lasing is obtained up to 130 K within the range of available input power. This upper temperature limit is set by detuning of the quantum well gain peak away from the cavity resonance. Note that the threshold powers given here are those of the 808 nm laser at the input of our cryostat. The actual threshold power is significantly lower when one accounts for the light reflected and scattered by the upper DBR and the small absorption path of the quantum well gain region, in addition to losses incurred by one lens, two windows, and one mirror.

\begin{figure}[h]
\includegraphics[width=3.3in]{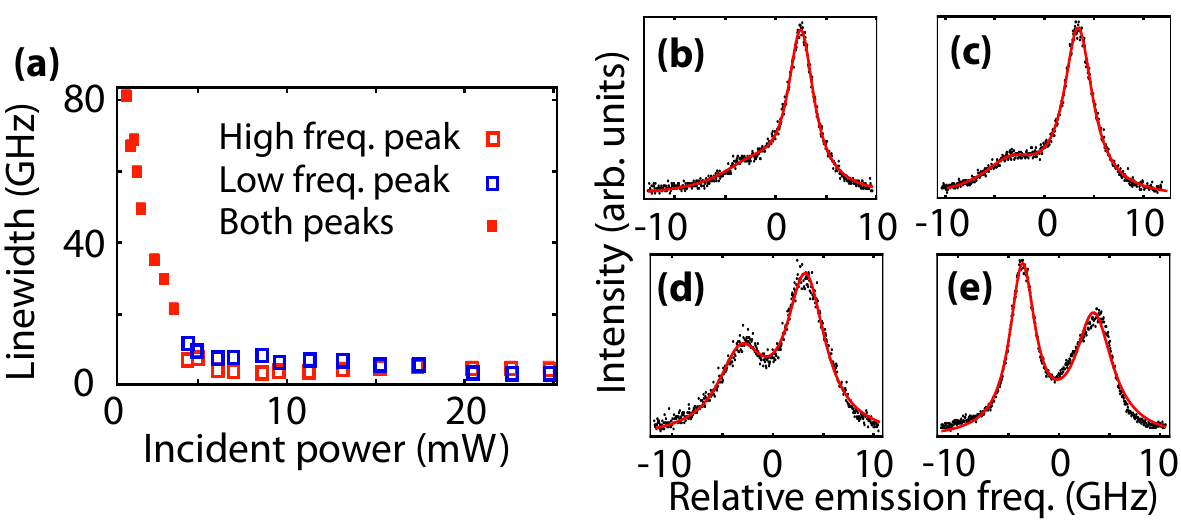}
\caption{\label{FP} (Color online) (a) Linewidth versus incident pump power for the same BH VCSEL as in Fig. 3. The solid red squares represent measurements with 20 GHz resolution (grating spectrometer). The open squares represent measurements with 35 MHz resolution of the fine-structure of the lowest order transverse mode of the BH VCSEL. (b-e) Spectra of the latter at 8.5 mW, 11.2 mW, 15.2 mW, and 22.6 mW incident pump power, respectively.}
\end{figure}

Figure 4 shows more details of the lasing properties for the 3 $\upmu$m BH VCSEL. In Fig. 4(a) the linewidth of the lowest order transverse mode [rightmost peak in bottom panel of Fig. 2(d)] is shown as a function of incident pump power. Measurements were performed with a grating spectrometer with a resolution 20 GHz and with a scanning Fabry-Perot interferometer with a resolution of 35 MHz. Pump powers up to 6 times the threshold power were used. The linewidth measurements far below threshold indicate that the 3 $\upmu$m cavity has a quality factor of $Q\approx$ 4000. This quality factor is that of the loaded cavity subject to significant absorption due to the three quantum wells. It cannot be compared directly to cavity quality factors measured with no gain medium \cite{akahane2003hpn} or a gain medium consisting of a single layer of InAs quantum dots.\cite{muller2006hqa, stoltz2005hqf} Substantial narrowing is measured above threshold, with a linewidth of about 3 GHz for the highest power. The high resolution spectra [Fig. 4(b-e)] also reveal a splitting of the lowest order transverse mode into two peaks of perpendicular polarization. As is seen in the panels in which the pump power is increased from 8.5 mW [Fig. 4(b)] to 22.6 mW [Fig. 4(e)], one of the peaks reaches threshold first.

In summary, we have presented an all-epitaxial BH VCSELs using quantum well gain regions. The buried approach provides the possibility of scaling the VCSEL to smaller sizes without suffering parasitic carrier losses. The sample described here was optimized for low-temperature operation in order to relax requirements on pump laser performance. However, next steps will aim at extending operation to room temperature and to electrical pumping. Future efforts will further focus on reducing the BH VCSEL to smaller diameters of order $\sim$1 $\upmu$m.

This work is supported by the Army Research Office under Grant No. W911NF-12-1-0046.

\end{document}